\documentclass[pre,floatfix,twocolumn,showpacs]{revtex4}

\usepackage{amsmath}
\usepackage{amsfonts}
\usepackage{mathrsfs}
\usepackage{epsfig}
\usepackage{epstopdf}
\usepackage{graphicx,color}
\usepackage{dcolumn}

\begin{document}
\title{`Hits' emerge through self-organized coordination in
collective response of free agents}

\author{Anindya S. Chakrabarti$^1$ 
and Sitabhra Sinha$^{2,3}$}

\affiliation{                    
 $^1$ Economics area, Indian Institute of Management, Vastrapur, 
Ahmedabad 380015, India.\\
$^2$ The Institute of Mathematical Sciences, CIT Campus, Taramani, Chennai
600113, India.\\
$^3$ National Institute of Advanced Studies, 
Indian Institute of Science Campus, Bangalore 560012, India.
}
,


\begin{abstract}
Individuals in free societies frequently exhibit striking coordination
when making independent decisions {\em en masse}. Examples include the
regular appearance of hit products or memes with substantially higher
popularity compared to their otherwise equivalent competitors, or
extreme polarization in public opinion. Such segregation of events
manifests as bimodality in the distribution of collective choices.
Here we quantify how apparently independent choices made by
individuals result in a significantly polarized but stable
distribution of success in the context of the box-office performance
of movies and show that it is an emergent feature of a system of
non-interacting agents who respond to sequentially arriving
signals. The aggregate response exhibits extreme variability amplifying
much smaller differences in individual cost of adoption. Due to
self-organization of the competitive landscape, most events elicit
only a muted response but a few stimulate widespread adoption,
emerging as ``hits''.
\end{abstract}

\maketitle

\section{Introduction}
Complex systems often exhibit nontrivial patterns
in the collective (macro) behavior arising from the individual
(micro) actions of many agents~\cite{Castellano09,Sen2013}.
Despite the high degree of variability in the characteristics of the
individuals comprising a group, it is sometimes possible to observe 
robust empirical regularities in the system
properties~\cite{Neda00,Challet00,Watts07}.
The existence of inequality in individual success, often
measured by wealth or popularity, is one such universal
feature~\cite{Sinha11}.
While agents differ in terms of individual attributes, these
can only partly explain the degree of this inequality~\cite{Salganik06}.
The outcomes often have a heavy-tailed distribution with
a much higher range of variability than that observed in the
intrinsic 
qualities.
Apart from the well-known
Pareto law for income (or wealth)~\cite{Pareto,Sinha06}, 
other examples include distributions of 
popularity for
books~\cite{Sornette04}, 
electoral candidates~\cite{Fortunato07}, online
content~\cite{Ratkiewicz10}, and 
scientific paradigms~\cite{Bornholdt11}.

\begin{figure}
\begin{center}
\includegraphics[width=0.99\linewidth]{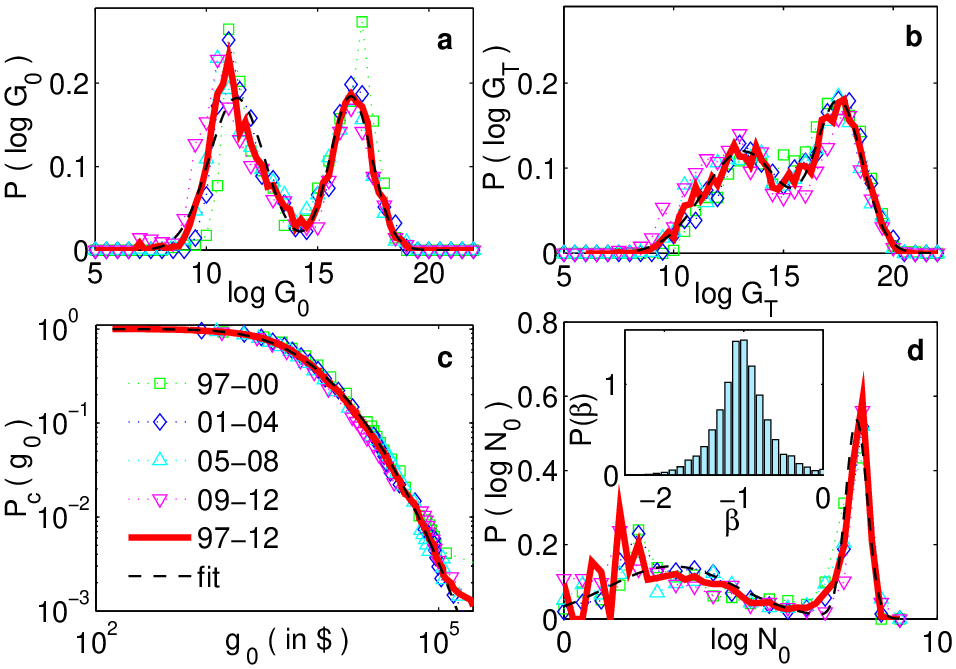}
\end{center}
\caption{(color online). Empirical demonstration of bimodality in
movie popularity measured in terms of
(a) opening income $G_O$ and
(b) total lifetime income $G_T$ of 
movies in theaters
over successive intervals from
1997-2012 (indicated by different symbols).
The data are fit by
superposition of two log-normal distributions (broken curve). The
cumulative distribution of the opening income per theater 
$g_O = G_O/N_O$ over the same period is shown in (c). A fit with
log-normal distribution is also indicated (broken curve). 
(d) The bimodal character of (a) and (b) can 
be connected to
the bimodality observed in the distribution of the number 
of opening theaters $N_O$ (i.e., the total number of theaters in which
a new movie is released). The inset shows the distribution of
exponents $\beta$ characterizing the power-law decay of
the weekly income per theater ($g_t \sim g_O~t^{\beta}$) for all
movies. Note that all logarithms are to base $e$.
}
\label{fig1}
\end{figure}

Another form of inequality may be observed in
distribution of outcomes
having a strongly {\em bimodal} character.
Here events are clearly segregated into two distinct classes,
e.g., corresponding
to successes and failures, respectively. 
While such distributions have
been reported in many different contexts, e.g., gene expression~\cite{Kaern05}, 
species abundance~\cite{Collins91, Hui12},
wealth of nations~\cite{Paap98},
electoral outcomes~\cite{Mayhew74, SinhaPan06}, etc., one of the most robust demonstrations
of bimodality is seen in the distribution of movie box-office
success~\cite{Pan10}. Here success is measured in terms of either the gross income $G_O$ at the opening weekend
or the total gross $G_T$ calculated over the lifetime (i.e., the entire duration that a movie
is shown) at theaters. Fig.~\ref{fig1}~(a-b) shows that 
both of these distributions 
constructed from publicly available data
for movies released in USA during the period 1997-2012 are
described well by a mixture of two log-normal distributions. Although
the movie industry has
changed considerably during this time, 
the characteristic 
properties of the distributions appear
to remain invariant over the successive intervals comprising the period.
The log-normal character can be explained by 
the probability of movie success being 
a product of many independent chance factors~\cite{Shockley57}, 
and is indeed
observed in the unimodal distribution of opening income per theater
$g_O$ [Fig.~\ref{fig1}~(c)]. 
However, the clear distinction of movies into two classes in terms of their box-office
performance (as indicated by the occurrence of two modes in the $G_O$ and $G_T$ distributions) 
does not appear to be simply related to their intrinsic
attributes~\cite{DeVany04}.
The fact that bimodality is manifested at the very beginning of a movie's
life also suggests that 
the extreme divergence of outcomes cannot be fully attributed to
social learning occurring over time as a result
of diffusion of information about
movie quality~\cite{Moretti11} (e.g., by word-of-mouth~\cite{Liu06}). We also emphasize that
the bimodal behavior is extremely robust and existed even before the
advent of social media, 
which plays a major role in word-of-mouth dynamics~\cite{Ishii2012}.
Thus,
while there have been theoretical attempts to
explain the emergence of bimodality by assuming specific forms of interactions between
agents~\cite{Watts02}, it is of interest to see if bimodal popularity distributions can arise 
without explicit agent-agent interactions.

In this paper, we present a model for understanding 
the collective response of a system of
agents to successive external shocks,
where the behavior of each agent
is the result of a decision process independent of other agents.
Even in the absence of explicit interaction among agents, the system
can exhibit remarkable coordination, characterized by the appearance
of a strong bimodality in its response.
For the specific example of box-office success, the bimodal nature of the gross income distributions
appears to be connected to the fact that movies usually open in either many or very few theaters.
Therefore, we focus 
on explaining the appearance of a bimodal
distribution for the number of theaters $N_O$ in 
which movies open [Fig.~\ref{fig1}~(d)].
Similar to how the observed invariant properties of financial markets
can be reproduced by agents interacting indirectly
through their response to a common
signal (price)~\cite{Vikram11}, our model comprises agents (theaters) that do not explicitly interact with each other but 
whose actions achieve coherence by the regular arrival
of a global stimulus, viz., new movies being introduced in the market.
By contrast, decoherence is induced by the uncertainty under which
each agent independently makes a decision on whether to switch to exhibiting the new movie or not.
We show that these competing effects can result in the
appearance of bimodality in the distributions of $N_O$, and consequently, $G_O$ and $G_T$, where the 
success of a particular movie cannot be simply connected to its perceived quality prior to release nor to its
actual performance on opening. 
Under a suitable approximation, we have analytically solved the model 
and obtained closed form expressions for the
peaks of the resulting multimodal distribution 
that 
match our numerical results. An important
implication of our study is that the box-office performance of a movie is crucially dependent on whether
it is released close in time to a highly successful one, which supports the popular wisdom that correctly timing the
opening of a movie determines its fate at box-office.

The paper is structured as follows.
In the next section, we discuss the empirical data on movie income
and its analysis in detail, while the model is introduced in
Sec.~\ref{model}. Section~\ref{results} describes the results, where
we also show the robustness of the bimodality obtained by looking at several
variants of the basic model. In addition, we provide
an analytical explanation for the emergence of bimodality in the model.
We conclude with a discussion of the implications of our findings in
Sec.~\ref{discussion}.

\section{Data analyzed}
\label{data}
\noindent {\em Data description.}
Income distributions are computed from publicly available data
(obtained from {\em The Movie Times} website~\cite{movietimes})
on box-office performance of movies released in the United
States of America over a span of 16 years (1997-2012). 
Gross income over all theaters within the USA are considered and the
data are inflation-adjusted with respect to 2010 as base year. 
To determine the time-invariance of the nature of income distribution,
the total time period has been
divided into four intervals, viz., 1997-2000, 2001-2004, 2005-2008, and
2009-2012. The total number of movies for which opening weekend gross
income $G_O$
data is available in
each of these intervals is 673, 1240, 1444, and 1226, respectively,
while total income $G_T$ (i.e., box-office receipts over the entire
period that a
movie was shown in theaters) is available for 1160, 1240, 1444, and 1226
movies in each of these intervals, respectively. 
Note that, a movie is associated with the
calendar year in which it was released in theaters within the USA.
Time-series of box-office income
has been obtained from {\em The Movie Times} site~\cite{movietimes} 
for a total of 4568 movies 
over the period July 1998 to July 2012.
To obtain opening weekend income per theater $g_O$, the gross opening income 
$G_O$ is divided by the number of movie
theaters $N_O$ in which the movie is released in its opening week.
\begin{table}[bp]
\caption{Values of log-normal distribution parameters for
different aggregate variables in the empirical data
estimated by maximum likelihood procedure.}
\begin{center}
\begin{tabular}{|l|c|c|c|c|c|c|}
\hline
\hline
Variable & Distribution type & $\alpha$ &$\mu_1$&$\mu_2$&$\sigma_1$&$\sigma_2$\\
\hline
$N_O$ & Bimodal & 0.61 & 2.91 & 7.84 & 1.72 & 0.29\\
\hline
$G_O$ & Bimodal & 0.57 & 11.36 & 16.49 & 1.24 & 0.94\\
\hline
$G_T$ & Bimodal & 0.54 & 13.16 & 17.55 & 1.80 & 1.05\\
\hline
$g_O$ & Unimodal &      & 8.72 &            & 1.02 &   \\
\hline
$N_{max}$ & Bimodal & 0.61 & 4.01 & 7.83 & 1.71 & 0.27\\
\hline
\end{tabular}
\label{table1}
\end{center}
\end{table}

\noindent {\em Fitting procedures and statistical tests.}
The aggregate variables $N_O$, $G_O$, and $G_T$ are fit with
bimodal log-normal distributions, i.e., a mixture of two log-normal
distributions with parameters $\mu_1$, $\sigma_1$ and $\mu_2$,
$\sigma_2$, that are weighted by factors $\alpha$ and $1-\alpha$,
respectively. The unimodal distribution of opening income per theater,
$g_O$, has been fit with a log-normal distribution having parameters
$\mu$ and $\sigma$.
The maximum likelihood estimates (MLE) of the parameters for the
empirical distributions of $N_O$,
$G_O$, $G_T$, and $g_O$ are shown in Table~\ref{table1}. 
Hartigan's dip test~\cite{Hartigan1985} for multimodality has been 
performed on the data for $N_O$, $G_O$, and $G_T$ and unimodality is 
rejected at $5\%$ significance level. By contrast, unimodality for 
the distribution of $g_O$ is not 
rejected by the test.
The time-series of movie income, $g_t$, has been fit to the general
form $g_t \sim g_O t^\beta$ by a regression procedure carried out over
all movies that were shown in theaters for at least 5 weeks.\\

\noindent {\em Robustness of empirical features.}
To see whether the qualitative features of the results of empirical
analysis are robust, we
have also looked at variables other than $N_O$, $G_O$, $G_T$ and $g_O$. For
example, if we consider instead of the opening number of theaters
$N_O$, the largest number of theaters $N_{max}$ that a movie is shown
simultaneously at any time following its release, its distribution
also shows a bimodal nature and can be fit by a 
superposition of two log-normal distributions 
Also, instead of considering only the opening income per theater
$g_O$, we have looked at the distribution of income per theater of a
movie at any given week following its release 
which 
is seen to be qualitatively similar to $g_O$
and can be fit by a unimodal log-normal distribution.


\section{The Model}
\label{model}
\begin{figure}
\begin{center}
\includegraphics[width=0.99\linewidth]{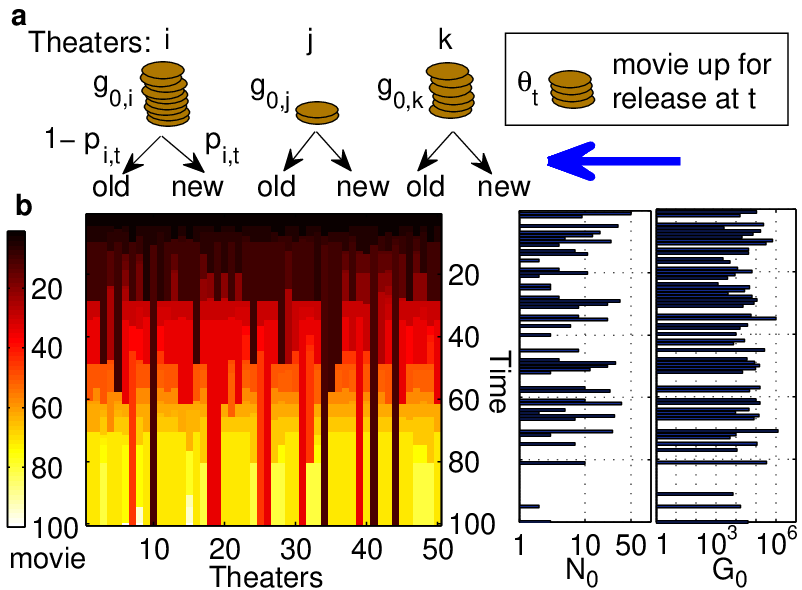}
\end{center}
\caption{(color online).
(a) Schematic diagram of the stochastic decision process of agents
(theaters $i$, $j$ and $k$) 
who can either continue with ``old'' (movie being shown) or switch to
``new'' (movie up for release) at any time instant $t$.
The probability that an agent $i$ will adopt the new movie, $p_{i,t}$,
depends on a comparison of the perceived performance of that
movie, $\theta_t$, to the actual performance of the movie being shown
(which is related to its opening income $g_{0,i}$).
(b) Time-evolution of a system comprising $N = 50$ agents (theaters), the state
of each agent at any time being the movie (colored
according to the time of release) that it is showing.
At every time instant, a new movie is available for release.
The variable performance of these movies are indicated 
in terms of the number of theaters
where they open ($N_O$) and their opening income ($G_O$).}
\label{fig2}
\end{figure}

We consider a system comprising $N$ agents ({theaters or
theater chains}) 
subjected to external stimuli (entry of new movies into the market),
that have to choose a
response, i.e., whether or not to adopt a new movie, displacing the one being
shown. 
At any time instant $t$, this decision depends on a comparison 
between the perceived
performance of the new movie and the actual
performance of the movie being shown at the theater
[Fig.~\ref{fig2}~(a)].
For simplicity, we assume that
a single new movie is up for release at each time instant $t$, 
thus allowing each movie to be identified by the
corresponding value of $t$. 
Allowing multiple movies to be 
released together does not qualitatively change the results.
The state of a theater at any time is indicated by the identity of the
movie it screens at that time [Fig.~\ref{fig2}~(b)].
The performance of a movie $t^{\prime}$ at time $t$
can be 
quantified by the estimated
income per theater, $g^t$, which is
related to its opening value $g_O^{t^{\prime}}$ by a scaling relation $g^t =
g_O^{t^{\prime}} (t - t^{\prime})^{\beta_s}$. 
This relation is partly
inspired by the empirical observation [Fig.~\ref{fig1}~(d), inset] that
the weekly income per theater for a movie decays as a power law
function of the number of weeks after its release, characterized by exponent
$\beta$~\cite{Pan10}. One
can also interpret $\beta_s$ as a subjective discount factor employed
by the agents to estimate the future income of a movie based on its
present income.
For simplicity, most results presented here are for $\beta_s = 0$. We
also show that other choices of $\beta_s$ yield qualitatively similar
results.
Note that the model does not assume any competition for audience 
between theaters showing the same movie (i.e., the demand
is perfectly inelastic in terms of theaters)
as the empirical data suggests that the income per theater of a movie
is relatively independent of the number of theaters in which it opens.

\noindent{\em Information available to agents.} 
As agents are exposed to similar information about
a movie that is up for release,
they can have a
common perception about its performance, measured as its predicted
opening income per theater, 
$\theta_t$.
This is chosen at each time step from a distribution that is identical
to that of $g_O^t$.
In fact, if the agents had perfect foresight, this prediction would be
identical to the actual opening income of the movie $g_O^t$, which
would have resulted in either a movie releasing in all theaters or
not being released in any theater. 
In general, however,
predictions are rarely accurate~\cite{DeVanyWalls_04} and the results
shown here are obtained for the case when the predicted income
$\theta_t$ is independent of the realized income $g_O^t$. 
We later show that the qualitative
behavior of the model is unchanged even when $\theta_t$ is correlated
with $g_O^t$.

\noindent{\em Dynamics of the adoption process.}
At any time $t$, an agent $i$ switches to the new movie if it decides
that this move will result in {a sufficiently high} net gain $z_{t} (i) = \theta_t -
g^{t} (i)$, {measured as the difference between the predicted 
income of the new movie up for release and the income of the currently 
running movie. As $\theta_t$ is log-normally distributed with
the $\mu$ and $\sigma$ of $g_O$ estimated from empirical data
(see Table \ref{table1}), we
normalize $z_t(i)$ by the mean of the distribution, viz.
${\rm exp} (\mu+\sigma^2/2)$. The action of switching (or not)} is implemented 
by representing the probability
of adopting the new movie as a hyperbolic response
function~\cite{Real77} for positive net gain $z_t(i)$:
\begin{equation}
p_{i,t}[z_t(i)] =
\begin{cases}
 \frac{z_{t}(i)}{C+z_{t}(i)}~& \text{for}~z_{t} (i) \geq 0, \\
               0 & \text{otherwise,}
\end{cases}
\label{eq:func1}
\end{equation}
where
parameter $C$ is the cost of adoption, incurred 
due to switching to a new movie.
Such a functional form allows us to model probabilistic decision making 
under uncertainty by the agents. 
At the
limit of extremely low adoption cost, i.e., $C \rightarrow 0$, we recover a more deterministic
switching behavior from Eq.~(\ref{eq:func1}), with the probability of adoption behaving as a step function as
it changes from 0 to 1 around $z=0$.
Eq.~(\ref{eq:func1})
allows us to
calculate the number of opening theaters $N_O$ for every new movie
[Fig.~\ref{fig2}~(b)].
To obtain the opening income $G_O$ of the movie over all theaters that
release it, $N_O$ is multiplied with the opening income per theater
that is chosen from the log-normal distribution of $g_O$ referred to
earlier [Fig.~\ref{fig1}~(c)].
The subsequent decay of income per theater follows the
empirical scaling relation with exponent $\beta$~\cite{Pan10}. The total
lifetime income of a movie $G_T$ is obtained by aggregating this
income for all theaters it is shown in, over the entire lifespan
(i.e., from the time it is released until it is displaced from all
theaters). 
The first few hundred time steps of each simulation realization were
considered to be transients and removed to avoid initial state
dependent effects.
\begin{figure}
\begin{center}
\includegraphics[width=0.99\linewidth]{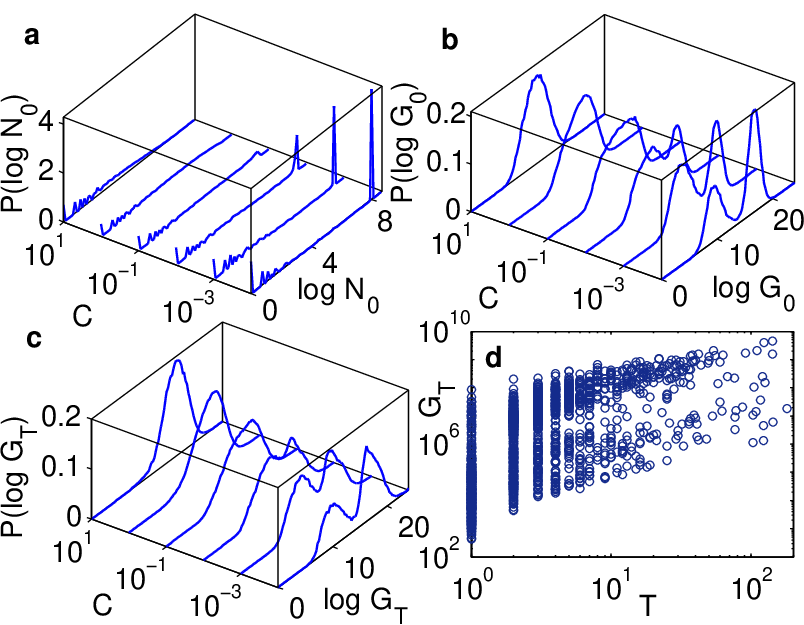}
\end{center}
\caption{(color online).
A bimodal distribution emerges from independent decisions of $N$ agents
(theaters).
Transition between bimodality and unimodality
with parametric variation of the cost of adoption $C$ is shown for the
distributions of (a) the number of opening theaters $N_O$, (b) 
opening income $G_O$ and (c) total lifetime income $G_T$ of 
movies. 
The results are obtained
by averaging over 
60 realizations 
with $N = 3000$ agents. 
(d) The total income $G_T$ earned by a movie as a function of its lifetime
$T$, i.e., the duration of its run at theaters, shows that for higher
values of $T$, the movies separate into two classes ($C=10^{-4}$).
}
\label{fig3}
\end{figure}

\section{Results}
\label{results}

\begin{figure}[tbp]
\begin{center}
\includegraphics[width=0.99\linewidth]{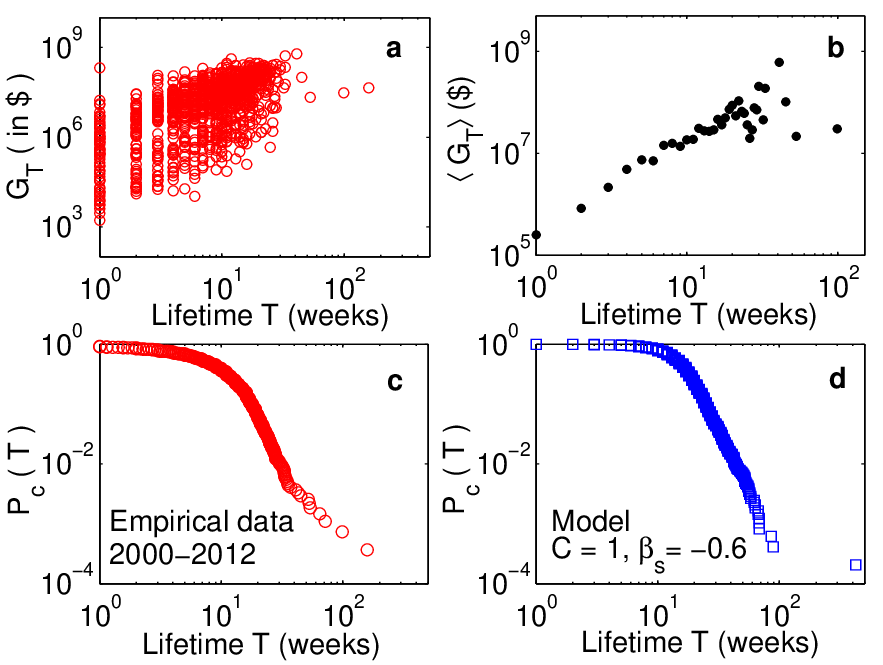}
\end{center}
\caption{(color online).
(a) Total lifetime income $G_T$ of movies released
during 2000-2008 shown as a function of
the number of weeks $T$ that they were shown in theaters.
(b) The average total gross $\langle G_T \rangle$
corresponding to each value of $T$. At large values of $T$, we observe
a divergence corresponding to a separation of the movies into two
classes.
(c) Complementary cumulative distribution of the lifetime $T$ of
movies, i.e., the duration of their run in theaters, for movies
released during 2000-2012. (d) The corresponding distribution
generated by the model system for $N$ = 3000 agents (theaters), cost
of adoption $C$ = 1, and subjective discount factor
$\beta_s = -0.6$. Note that the shape of the distribution obtained
from the model
can be varied to an extent by changing the parameters $C$ (that shifts
the distribution along the horizontal axis) and $\beta_s$ (which
alters the slope).}
\label{fig5}
\end{figure}

\noindent{\em Reproducing the bimodal distribution of movie income.}
As seen from Fig.~\ref{fig3}, the 
system of $N$ independent agents
self-organize in the limit of low $C$ to generate a bimodal
distribution in their collective response. A new movie is either
adopted by a majority [corresponding to the upper mode 
of the $N_O$ distribution 
shown in Fig.~\ref{fig3}~(a)] or a small fraction [lower mode] of the
total number of theaters. This translates into bimodal distributions
in the opening income $G_O$ and total lifetime income $G_T$
[Fig.~\ref{fig3}~(b-c)], which qualitatively resemble the
corresponding empirically obtained distributions (Fig.~\ref{fig1}).
To emphasize that bimodality in total income $G_T$ is a
consequence of the bimodal nature of the opening income, we show $G_T$ as 
a function of the lifetime $T$ in Fig.~\ref{fig3}~(d). We observe a
bifurcation in $G_T$ at higher values of $T$ indicating that movies
having the same lifetime can have very different total income,
a feature that is seen in empirical data [Fig.~\ref{fig5}~(a-b), see also
Ref.~\cite{Sinha04}]. 
Thus, our results suggest that the nature of box-office income
distributions for movies can be understood as an outcome of the
bimodal character of the distribution for the number of theaters 
that release a movie coupled with the unimodal log-normal
distribution for the income per theater.

A verification of our model results with empirical data is
provided by a comparison of the corresponding distributions of the
lifetime of movies, i.e., the duration of their run in theaters.
Fig.~\ref{fig5}~(c-d) shows that the two distributions are
qualitatively similar. The shape of the lifetime distribution for the
model can be varied to an extent by changing the cost of adoption $C$
and the subjective discount factor $\beta_s$.

\noindent{\em Transition to unimodality with increasing adoption
cost.}
As the cost of adoption $C$ is increased, the two modes approach each
other until, at a large enough value of $C$, a
transition to unimodal distribution for the quantities is observed
[Fig.~\ref{fig3}~(a-c)].
With increasing $C$, theaters are less likely to switch to a new
movie, so that the time-interval between two consecutive movie
releases at a theater becomes extremely long. This weakens
temporal correlations between the performance of movies being shown
and that expected from new movies up for release. Thus, the decision to release
each new movie eventually becomes an independent stochastic event
described by an unimodal distribution.

\begin{figure}[tbp]
\begin{center}
\includegraphics[width=0.99\linewidth]{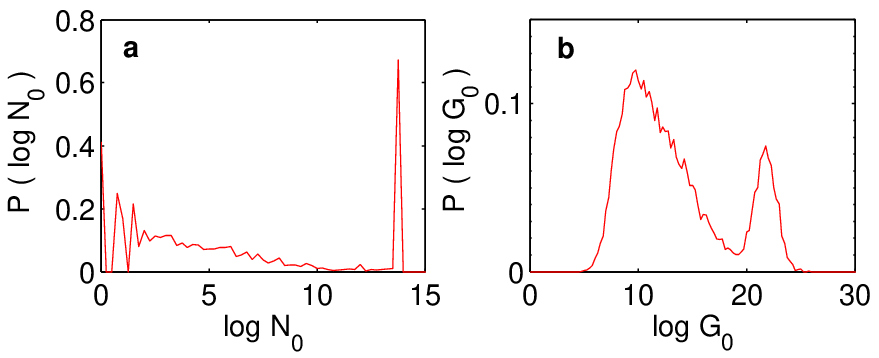}
\end{center}
\caption{(color online). 
Robustness of the model results with respect to variation in
system size $N$, viz., increasing the number of agents to $N=10^6$.
The results are obtained by simulating the system with cost of
adoption $C = 10^{-4}$ for $500$ iterations and
averaging over $60$ realizations. The bimodal nature of
the distributions of (a) the number
of opening theaters $N_O$ and (b) opening income $G_O$ of movies
is evident.}
\label{system_size_independence}
\end{figure}

\noindent{\em Robustness of bimodality.}
For most simulations, we have chosen $N = 3000$, which
accords with the maximum number of theaters in the empirical 
data. However, to verify that our results are not sensitively
system-size dependent, we have checked that qualitatively similar
behavior is observed for $N$ up to $10^6$ (Fig.~\ref{system_size_independence}).

\begin{figure}[tbp]
\begin{center}
\includegraphics[width=0.99\linewidth]{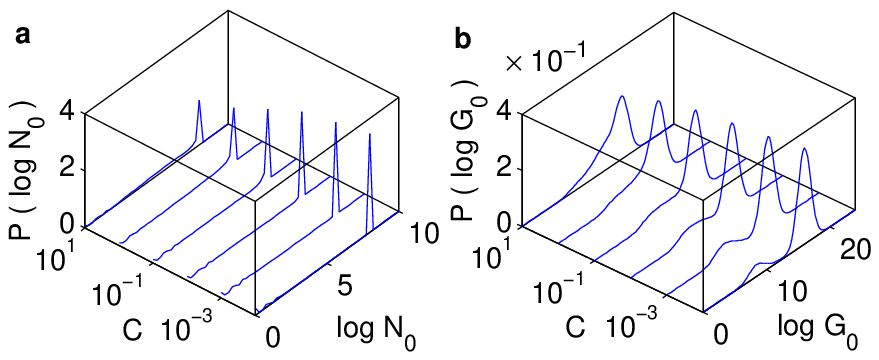}
\end{center}
\caption{(color online). 
Robustness of the model results with respect to a different
choice of the subjective discount factor, viz., $\beta_s=-1$. 
While the peaks at the lower value are smaller than the case of
$\beta_s = 0$ for both (a) the number of opening theaters $N_O$ and
(b) opening income $G_O$, it can be observed that as the cost of
adoption, $C$, is decreased, the nature of the distribution changes from
unimodal to bimodal. 
Results are shown for $N$ =3000 agents for $10^4$ iterations and
averaged over 60 realizations.}
\label{beta1}
\end{figure}

While for most results reported here the subjective discount factor 
$\beta_s = 0$, we have verified that 
the results are qualitatively unchanged if
$\beta_s$ has a value different from $0$. Fig.~\ref{beta1}
shows that even if $\beta_s = -1$, a transition from
unimodality to bimodality occurs as seen earlier for $\beta_s = 0$,
when the cost of adoption $C$ is decreased.

\begin{figure}[tbp]
\begin{center}
\includegraphics[width=0.95\linewidth]{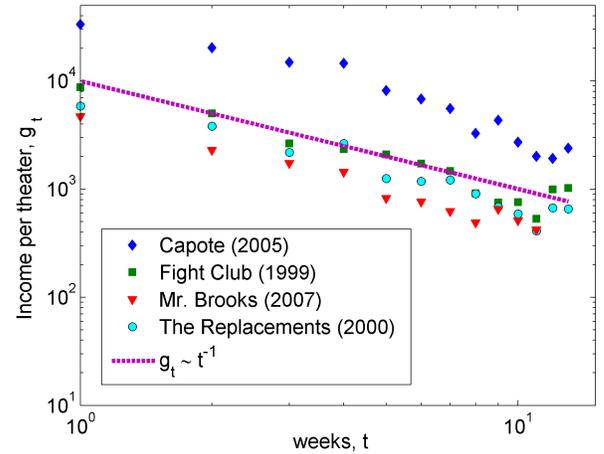}
\end{center}
\caption{(color online). The time-evolution of the income per theater $g$ of four
movies that were released in theaters at various times during the
period investigated here. The decay of $g_t$ with $t$ approximately
fits a power-law form. The broken line corresponding to $g_t\sim
t^{-1}$
is shown for visual reference.
}
\label{beta_1_empirical}
\end{figure}

The empirical data shows that the income per theater of all movies
decay with time having an approximately power-law form $g_t \sim g_O
t^{\beta}$ (Fig.~\ref{beta_1_empirical}). 
The value of the exponent $\beta \approx -1$ on average
(corresponding to the broken line in Fig.~\ref{beta_1_empirical}),
which governs how the income per theater changes over time.
This motivated our choice of $\beta = -1$ in the basic model. 
\begin{figure}[tbp]
\begin{center}
\includegraphics[width=0.99\linewidth]{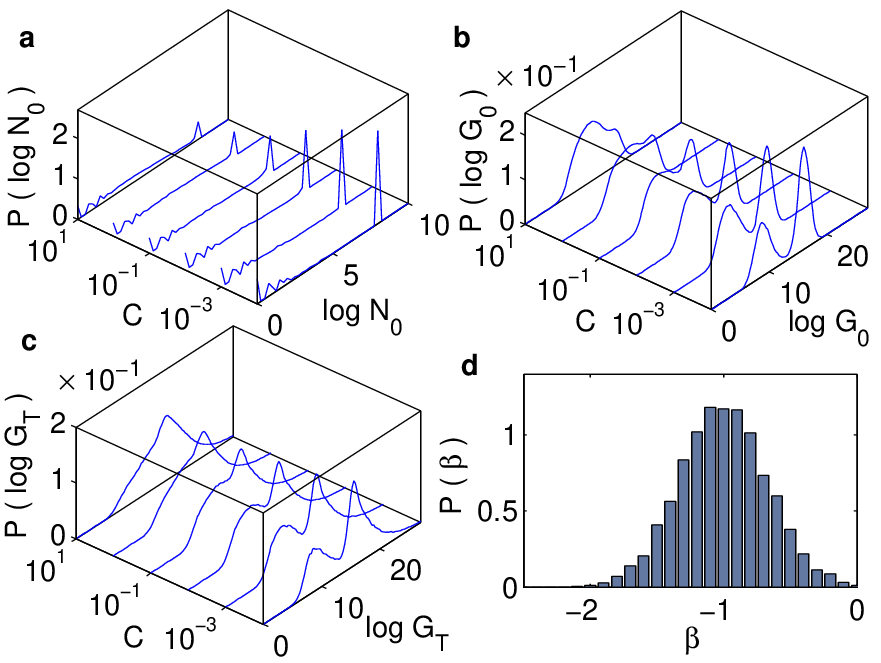}
\end{center}
\caption{(color online). Robustness of the model results with respect to heterogeneity
in the nature of the temporal decay of the income per theater of
different movies, viz., their decay exponents $\beta$ being distributed
approximately as the corresponding empirical distribution shown in
Fig.~\ref{fig1}~(d, inset).
Transition between bimodality and unimodality
with parametric
variation of the cost of adoption $C$ is shown for the distributions
of (a) the number of opening theaters $N_O$, (b) opening income
$G_O$ and (c) total lifetime income $G_T$ of movies. 
The results are obtained by simulating a system with $N = 3000$
agents for $10^4$ iterations and
averaging over $60$ realizations. The distribution of $\beta$ for
different movies for a particular simulation realization is shown in
(d). The values are generated from a normal
distribution with the same mean ($\mu = -1$) and standard deviation ($\sigma
= 0.33$) as the empirical
distribution of $\beta$.
}
\label{beta_distributed}
\end{figure}
Instead of all movies having exactly identical form of decay in
the time-evolution of their income per theater as in the basic
model, we can consider that
different movies are characterized by different values of the exponent
$\beta$. 
In particular, we choose the values of $\beta$ from a distribution
that approximates the empirical distribution of $\beta$ shown in
Fig.~\ref{fig1}~(d, inset). Fig.~\ref{beta_distributed} shows that the results
of the simulations of this variant model are qualitatively similar to
that of the basic model, including the transition from unimodality to
bimodality.

\begin{figure}[tbp]
\begin{center}
\includegraphics[width=0.99\linewidth]{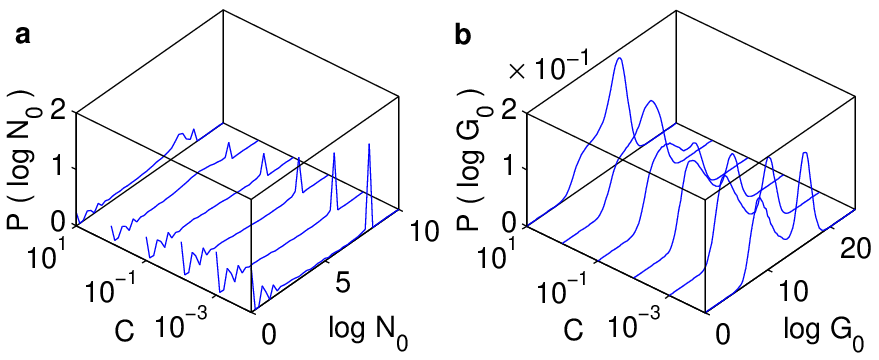}
\end{center}
\caption{(color online). Robustness of the model results with respect to use of
a different functional form for the adoption rule, viz., having a
sigmoidal character. Transition between bimodality and unimodality
with parametric
variation of the cost of adoption $C$ is shown for the distributions
of (a) the number of opening theaters $N_O$ and (b) opening income
$G_O$, when Eqn.~(\ref{Eqn:strat_2}) is used for the
functional form representing the probability of adopting the new
movie. The results are obtained by simulating a system with $N = 3000$
agents for $10^4$ iterations and
averaging over $60$ realizations. 
}
\label{strategy2}
\end{figure}

\begin{figure}[tbp]
\begin{center}
\includegraphics[width=0.99\linewidth]{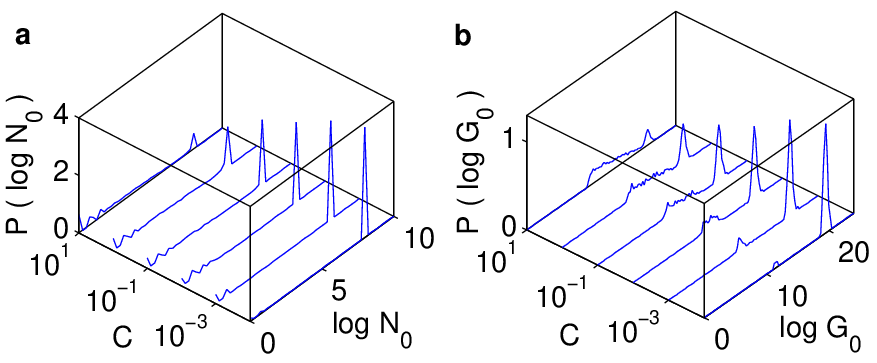}
\end{center}
\caption{(color online). 
Robustness of the model results when the agents (theaters)
can exactly predict their income from a new movie up for release,
i.e., $\theta = g_O$. 
The distributions of (a) the number of opening
theaters $N_O$ and (b) opening income $G_O$ are unimodal
when the cost of adoption $C$ is low, as in this situation, a
movie will either be adopted by all theaters or none at all.
With increasing $C$, a distinct bimodal nature emerges in
the distributions.
Results are shown for $N$ =3000 agents for $10^4$ iterations and
averaged over 60 realizations.}
\label{exact_prediction}
\end{figure}

We have also verified that considering income aggregated over successive
periods (instead of only the opening income)
do not qualitatively change the results reported here.
The model also shows very similar behavior if, instead of Eq.~(1), we use
other more complicated functional forms for the probabilistic choice functions, e.g.,
\begin{equation}
p_{i,t}= \frac{1}{2} + \frac{z_t(i)}{2 \sqrt{C+z_t(i)^2}},
\label{Eqn:strat_2}
\end{equation}
which has a sigmoidal profile (Fig.~\ref{strategy2}). 
In addition, 
we have considered a variant model where the agents can make
perfect
prediction about the performance of a movie up for release so that
$\theta_t =
g_{O}^t$. 
Results are qualitatively similar to the basic model and bimodality is
see over a range of values of the cost parameter $C$
(Fig.~\ref{exact_prediction}).

\begin{figure}
\begin{center}
\includegraphics[width=0.99\linewidth]{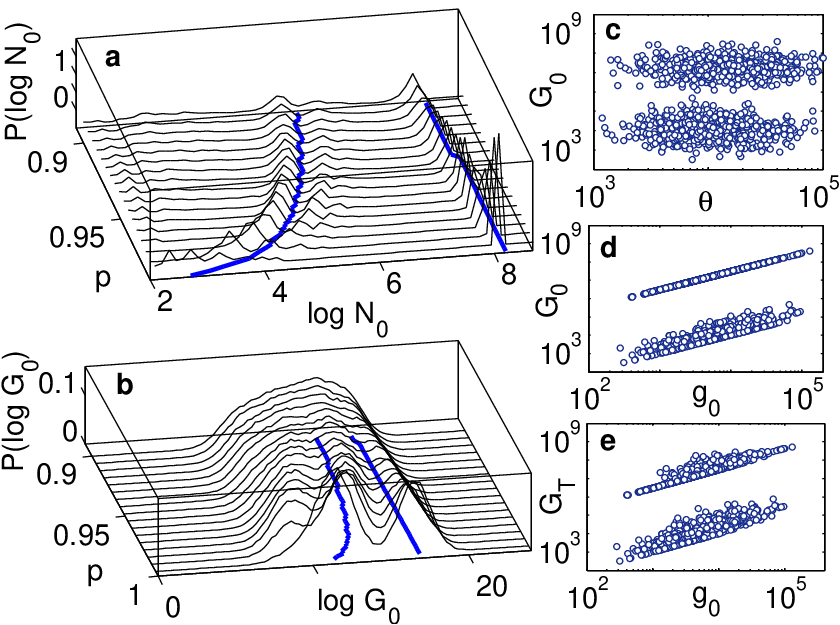}
\end{center}
\caption{(color online). Explaining the emergence of 
bimodal distribution in the limit of small cost of adoption ($C
\rightarrow 0$).
The appearance of bimodality
with parametric variation of the probability of adoption $p$ is shown for the
distributions of (a) the number of opening theaters $N_O$ and (b) 
opening income $G_O$. 
As $p \rightarrow 1$, the approximation to the $C
\rightarrow 0$ limit becomes more accurate.
The results are obtained
by averaging over 
60 realizations
with $N = 3000$ agents.
The pair of thick lines in each figure indicate the theoretically 
predicted modes of the distributions (see text).
(c-e) The variations of opening income $G_O$ and total lifetime income
$G_T$ of a movie as functions of the perceived performance $\theta$
and the actual performance (i.e., income per theater) $g_O$ shows that
neither $\theta$ nor $g_O$ completely determine $G_O$ or $G_T$ ($p =
0.9995$).}
\label{fig4}
\end{figure}

\noindent{\em Analytical explanation of the emergence of bimodality.}
To understand the appearance of multiple peaks in the distribution of
collective response in the limit of low cost of adoption, we observe that
the system dynamics is characterized by two competing effects: (a) the
stochastic decision process of the individual theaters tend to
increasingly decorrelate their states, while (b) the occasional
appearance of movies having high $\theta$, that are perceived by 
the agents to be potential box-office successes, induces high level of
coordination in response as a majority of agents switches to a common
state. 
This phenomenon of gradual divergence in agent states
interrupted by sporadic ``reset'' events that largely synchronize the system
allows us to use the
following simplification of the model for an analytical
explanation. As $C \rightarrow 0$, we can approximate 
Eq.~(\ref{eq:func1}) by 
$p_{i,t} = p$ for $z_t (i) \geq 0$, else $p_{i,t} = 0$, which 
becomes accurate in the limit $p\rightarrow 1$.
Thus, when a reset event occurs, the decision of each agent is
a Bernoulli trial with probability $p$, so that the number of theaters 
that adopt
the new movie follows a binomial distribution with mean $Np$ and
variance $Np(1-p)$. In the limit $p \rightarrow 1$ the variance
becomes negligibly small and the distribution
can be effectively replaced by its mean. This will correspond to a peak at
$N_O^u = Np$, i.e., the higher mode.

A movie that immediately follows a reset event can result in
different responses from the agents depending on the value of $\theta$
associated with it. If this is larger than $g^t$ of all theaters, it
is yet another reset event, the response to which is the same as above.
However, if $\theta$ has a lower value that is nevertheless large enough to
cause those theaters [$\simeq N(1-p)$] that had not switched in the 
previous reset event to adopt the new movie with probability $p$, we
obtain another peak at $N_O^l = Np(1-p)$. This corresponds to the
lower mode of the distribution. As seen from Fig.~\ref{fig4}~(a), the
two peaks of $N_O$ distribution are accurately reproduced by $N_O^u$
and $N_O^l$.
In principle, the above argument can
be extended to show that a series of peaks at successively smaller
values of $N_O$ can exist at $Np(1-p)^2$, $Np(1-p)^3$, etc., but these
will not be observed for the system size we consider here.
The bimodal log-normal distribution of opening income $G_O$
results from a 
convolution of the multi-peaked distribution for $N_O$ with 
the log-normal distribution for $g_O$ (having
parameters $\mu, \sigma$). 
The two modes of this
distribution are calculated as $G_O^{u,l} = \exp({\mu + \log
N_O^{u,l}})$, which matches remarkably well with the numerical
simulations of the model [Fig.~\ref{fig4}~(b)].

%
While the individual behavior of agents are obviously dependent on the
intrinsic properties (such as $\theta$) associated with specific
stimuli,
the collective behavior of the system cannot be reduced to a simple
threshold-like response to external signals.
Fig.~\ref{fig4}~(c) shows that the opening income of different movies,
which are segregated into two distinct clusters, are not simply
determined by their perceived performance $\theta$, as 
one can find movies belonging to either cluster for
any value of this quantity.
Given that $\theta$ is only a prediction of the opening
performance of a movie by the agents, and it need not coincide with reality, 
one may
argue that the actual performance, i.e., the opening income per
theater $g_O$, will be the key factor determining the aggregate income
of the movie.
However, Fig.~\ref{fig4}~(d-e) show that neither the opening income nor the
total lifetime income (both of which show clear separation into two
clusters) can be explained as a simple function of the actual opening 
performance of the movie at a theater.

\section{Discussion}
\label{discussion}

Our results explain box-office success as an
outcome of competition between movies, where a new movie seeks
to open at as many theaters as possible by displacing the older ones.
Using an ecological analogy, a movie with high perceived performance
invades and occupies a large number of niches until it is displaced
later by a strong competitor. Thus,
highly successful movies rarely coexist.
This also implies that the response to a movie can be very different
depending on whether or not it is released close to a reset event,
i.e., the appearance of a highly successful movie (``blockbuster''). 
Therefore, our model provides explicit theoretical support to popular
wisdom that timing the release of a movie correctly is a
key determinant of its success at the box-office~\cite{Krider98}.

We also note that the knowledge of the time elapsed from the last blockbuster may
not by itself lead to a successful strategy for optimally timing the
release of a movie.
If the entry of a new movie is delayed to increase
the time interval from the previous reset event so as to increase its chance of doing well at the box office, 
a competing movie released before it may become a ``hit'' and thereby
prevent its success. 
Thus, there is a trade-off between waiting for as long as possible after the 
last successful movie but not so long as to get beaten by a competitor.
The critical importance of the launch time holds not only for
movies, but also for many other short life-cycle products such as music,
video games, etc., whose opening revenues very often decide their 
eventual sales~\cite{Friedman04}. In fact, empirical data on movies
show that for the dominant majority, the highest gross earning 
over all theaters they are shown in
occurs on the
opening weekend, followed by an exponential decay in
income~\cite{Pan10}. In extremely few cases does a movie become more
successful over time with its income exhibiting an increasing trend,
eventually reaching a peak before again declining exponentially. 
To explain such rare ``sleeper hits'' [e.g., the movie {\em My Big Fat
Greek Wedding} (2002) that achieved its highest gross around 20 weeks after
its release], one may need to consider 
how agents can directly influence each other.
This suggests that models for generating bimodal distributions that
incorporate explicit interactions between
agents such as in Ref.~\cite{Watts02} could complement the one presented here
where an effective external field guides the actions of the agents who
otherwise do not communicate.

To conclude, we have shown that extreme variability in response,
characterized by a bimodal distribution, can arise in a system even in
the absence of explicit interactions between its components.
The observed inequality of
outcomes cannot be explained solely on the basis of
variations in the intrinsic quality of signals driving the system.
For a quantitative validation of the model 
we have used the explicit example of movie box-office performance
whose bimodal distribution has been established empirically.
The log-normal nature of the distribution of income per theater
suggests that the underlying mechanism involves sequential stochastic
processes.
Our analysis reveals that stochastic decisions on the basis of
comparing effects of the preceding choice and the estimated impact of
the upcoming one gives rise to a surprising degree of coordination.
The presence of bimodality in the absence of explicit interactions
in several social and biological systems 
suggests other possible applications of the theoretical approach
presented here.
Apart from bimodality, our model shows that more general multimodal
distributions are possible in principle and empirical verification
of this in natural and social systems will be an exciting development.

\begin{acknowledgements}
We thank Alex Hansen, Gautam I. Menon, 
Shakti N. Menon, and Rajeev Singh for helpful
discussions. This work was supported in
part by IMSc Econophysics Project (XII Plan) funded by
the Department of Atomic Energy, Government of India.
\end{acknowledgements}


\begin{thebibliography}{10}
\bibitem{Castellano09}
C. Castellano, S. Fortunato and V. Loreto, Rev. Mod. Phys. {\bf 81}, 591 (2009).

\bibitem{Sen2013}
P. Sen and B.~K. Chakrabarti, {\em Sociophysics: An Introduction}
(Oxford Univ. Press, Oxford, 2013).

\bibitem{Neda00}
Z. N\'{e}da, E. Ravasz, Y. Brechet, T. Vicsek and A.-L. Barab\'{a}si, Nature (Lond.) {\bf 403}, 849 (2000).

\bibitem{Challet00}
D. Challet, M. Marsili, and R. Zecchina, Phys. Rev. Lett. {\bf 84}, 1824 (2000).

\bibitem{Watts07}
D.~J. Watts, Nature (Lond.) {\bf 445}, 489 (2007).


\bibitem{Sinha11} 
S. Sinha, A. Chatterjee, A. Chakraborti and B.~K. Chakrabarti, 
{\em Econophysics: An Introduction} (Wiley-VCH, Weinheim, 2011).

\bibitem{Salganik06}
M.~J. Salganik, P.~S. Dodds and D.~J. Watts, Science {\bf 311}, 854 (2006).

\bibitem{Pareto}
V.~M. Yakovenko and J.~B. Rosser,  Rev. Mod. Phys. {\bf 81}, 1703 (2009).

\bibitem{Sinha06}
A. Chatterjee, S. Sinha and B.~K. Chakrabarti, Curr. Sci. {\bf 92}, 1383 (2007).

\bibitem{Sornette04}
D. Sornette, F. Desch\^{a}tres, T. Gilbert and Y. Ageon,
Phys. Rev. Lett. {\bf 93}, 228701 (2004).

\bibitem{Fortunato07}
S. Fortunato and C. Castellano, Phys. Rev. Lett. {\bf 99}, 138701 (2007).

\bibitem{Ratkiewicz10}
J. Ratkiewicz, S. Fortunato, A. Flammini, F. Menczer
and A. Vespignani, Phys. Rev. Lett. {\bf 105}, 158701 (2010).

\bibitem{Bornholdt11}
S. Bornholdt, M.~H. Jensen and K. Sneppen, Phys. Rev.
Lett. {\bf 106}, 058701 (2011).

\bibitem{Kaern05}
M. Kaern, T.~C. Elston, W.~J. Blake and J.~J. Collins,
Nature Rev. Genet. {\bf 6}, 451 (2005).

\bibitem{Collins91}
S. L. Collins and S. M. Glenn, Ecology {\bf 72}, 654 (1991).

\bibitem{Hui12}
C. Hui, Community Ecology {\bf 13}, 30 (2012).

\bibitem{Paap98}
R. Paap and H.~K. van Dijk, Eur. Econ. Rev. {\bf 42}, 1269 (1998).

\bibitem{Mayhew74}
D.~R. Mayhew, Polity {\bf 6}, 295 (1974).

\bibitem{SinhaPan06} S. Sinha and R.~K. Pan, 
in {\em Econophysics and Sociophysics}, eds. B.~K. Chakrabarti {\em et al.}
(Wiley-VCH, Weinheim, 2006), p. 417.

\bibitem{Pan10} R.~K. Pan and S. Sinha, 
New J. Phys. {\bf 12}, 115004 (2010).


%

\bibitem{Shockley57}
W. Shockley, Proc. IRE {\bf 45}, 279 (1957).

\bibitem{DeVany04}
A. De Vany, {\em Hollywood Economics} (Routledge, London, 2004).

\bibitem{Moretti11}
E. Moretti, Rev. Econ. Stud. {\bf 78}, 356 (2011).

\bibitem{Liu06}
Y. Liu, J. Marketing {\bf 70}, 74 (2006).

\bibitem{Ishii2012}
A. Ishii, H. Arakaki, N. Matsuda, S. Umemura, T. Urushidani, N. Yamagata and N. Yoshida,
New J. Phys. {\bf 14}, 063018 (2012).

\bibitem{Watts02}
D.~J. Watts, Proc. Natl. Acad. Sci. USA {\bf 99}, 5766 (2002).

\bibitem{Vikram11}
S.~V. Vikram and S. Sinha, Phys. Rev. E {\bf 83}, 016101 (2011).

\bibitem{movietimes}
\url{http://www.the-movie-times.com}

\bibitem{Hartigan1985}
J.~A. Hartigan and P.~M. Hartigan, Annals Stat. {\bf 13}, 70 (1985). 

\bibitem{DeVanyWalls_04}
A.~S. De Vany and W.~D. Walls, J. Econ. Dyn. Contr. {\bf 28}, 1035 (2004).

\bibitem{Real77}
L.~A. Real, Am. Nat. {\bf 111}, 289 (1977).

\bibitem{Sinha04}
S. Sinha and S. Raghavendra, 
Eur. Phys. J. B {\bf 42}, 293 (2004).

\bibitem{Krider98}
R.~E. Krider and C.~B. Weinberg, J. Marketing Res. {\bf 35}, 1 (1998).


\bibitem{Friedman04}
R.~G. Friedman, in {\em The Movie Business Book}, ed. J.~E. Squire 
(3rd ed., Fireside, New York, 2004), p. 282.


\end{thebibliography}
\end{document}